\documentclass[preprint,showpacs,preprintnumbers,amsmath,amssymb]{revtex4}

\usepackage{epsfig} 
\usepackage{graphicx}
\usepackage{dcolumn}
\usepackage{bm}
\usepackage{wasysym}

\newcommand{\be}{\begin{equation}}
\newcommand{\ee}{\end{equation}}
\newcommand{\bea}{\begin{eqnarray}}
\newcommand{\eea}{\end{eqnarray}}

\newcommand{\mr}{\mathrm}

\def\llnnp{e^+\nu_e\mu^+\nu_\mu}
\def\llnnm{e^-\bar{\nu}_e\mu^-\bar{\nu}_\mu}
\def\wpp{W^+W^+}
\def\wmm{W^-W^-}

\begin{document}

\preprint{KA-TP-09-2009}

\title{Next-to-leading order QCD corrections to \\ $\boldsymbol{W^+ W^+
    jj}$ and $\boldsymbol{W^- W^- jj}$  production via weak-boson fusion}

\author{B.\ J\"ager}
\affiliation{
Institut f\"ur Theoretische Physik und Astrophysik, Universit\"at W\"urzburg, \\
97074~W\"urzburg, Germany
}

\author{C.\ Oleari}
\affiliation{
Universit\`a di Milano-Bicocca and INFN Sezione di Milano-Bicocca, \\
        20126 Milano, Italy
}

\author{D.\ Zeppenfeld}
\affiliation{
Institut f\"ur Theoretische Physik, \\
Universit\"at Karlsruhe, KIT, \\
76128 Karlsruhe, Germany
}


\begin{abstract}
We present a next-to-leading order QCD calculation for $\llnnp jj$ and 
$\llnnm jj$ production
via weak boson fusion at a hadron collider in the form of a fully-flexible
parton-level Monte Carlo program, which allows for the calculation of
experimentally accessible observables within realistic selection cuts. The
QCD corrections to the integrated cross sections are found to be modest,
while the shapes of some kinematical distributions change appreciably compared to leading order. The residual scale uncertainties of the next-to-leading order
results are at the few-percent level.
\end{abstract}

\maketitle



\section{\label{sec:intro} Introduction}
Weak-boson fusion (WBF) reactions have been identified as a promising means for
gaining a thorough understanding of the electroweak gauge boson sector of the
Standard Model (SM) and extensions thereof.  WBF is considered as a possible
discovery channel for the Higgs
boson~\cite{RichterWas:1999sa,Kersevan:2002vu,Kinnunen:1999ak, 
Rainwater:1999sd,Kauer:2000hi}
and a suitable tool for a later determination of its couplings to gauge bosons
and fermions~\cite{Zeppenfeld:2000td,Belyaev:2002ua,Duhrssen:2004cv} and of
its CP properties~\cite{Hankele:2006ja,Hankele:2006ma}. As WBF reactions are
sensitive to the mechanism of electroweak symmetry breaking {\em per se},
they could help to spot signatures of physics beyond the SM such as strong
interactions in the electroweak
sector~\cite{Chanowitz:1985hj,Chanowitz:1988ft,Vega:1989tt,Dicus:1991im,Dicus:1991fk,Bagger:1993zf,Bagger:1995mk,Englert:2008tn}. Furthermore, WBF processes with
like-sign dilepton final states are an important background to various new physics scenarios.

In order to unambiguously identify new physics effects, precise predictions for signal
and background processes are crucial, including estimates of the theoretical
uncertainties. In the context of perturbation theory, the required accuracy
and a first assessment of the uncertainties associated with the truncation of
the perturbative expansion can only be achieved by the calculation of
next-to-leading order (NLO) QCD corrections for experimentally accessible
observables. NLO-QCD corrections have therefore been provided for several
WBF-induced processes within the SM, including electroweak $Hjj$, $Hjjj$,
$W^\pm jj$, $Zjj$, $W^+W^- jj$, $ZZjj$, and $W^\pm Z jj$
production at hadron colliders~\cite{Figy:2003nv,Berger:2004pca,Figy:2007kv,
Oleari:2003tc,Jager:2006zc,Jager:2006cp,Bozzi:2007ur}. QCD corrections have
also been determined for selected WBF processes in new physics scenarios,
such as~\cite{Konar:2006qx,Englert:2008wp}.  
More recently, the electroweak corrections~\cite{Ciccolini:2007ec} to the
Higgs boson signal in WBF have been computed, together with parts of the
two-loop QCD corrections~\cite{Harlander:2008xn} and the interference with
QCD-induced $Hjj$
production~\cite{Andersen:2007mp,Bredenstein:2008tm}. Supersymmetric
corrections to $pp\to Hjj$ have been presented in the Minimal Supersymmetric
Standard Model (MSSM)~\cite{Hollik:2008xn}.

In this work, we focus on $\llnnp jj$  production via WBF at a
hadron collider and present results for the CERN Large Hadron Collider
(LHC). We develop a flexible Monte Carlo program that allows for the
calculation of cross sections and kinematical distributions within arbitrary
selection cuts at NLO-QCD accuracy. We discuss the most important theoretical
uncertainties and quantify the impact of radiative corrections on a few
representative distributions.

Results for the $\llnnm jj$ channel can simply be obtained by charge
conjugation and parity reversal. For example, if one is interested in $pp \to \llnnm jj$ at the
LHC, it is sufficient to run our code for the process $\bar{p}\bar{p} \to
\llnnp jj$ with a centre-of-mass energy of 14~TeV, treat the final-state
charged leptons as if they were negatively charged, and reverse momentum directions when considering parity-odd distributions.  For this reason, in the
rest of the paper we will discuss only the $\llnnp jj$ process.

We start with a brief overview of the technical pre-requisites of the
calculation in Sec.~\ref{sec:tech}. Section~\ref{sec:num} contains our
numerical results. We conclude in Sec.~\ref{sec:conc}.


\section{\label{sec:tech} Technical Pre-requisites}
The calculation of NLO-QCD corrections to $pp\to \llnnp jj$ via WBF proceeds along the same lines as our earlier work on WBF $W^+W^- jj$, $ZZjj$, and $W^\pm Z jj$ production  in $pp$ collisions. The methods developed in Refs.~\cite{Jager:2006zc,Jager:2006cp,Bozzi:2007ur} can therefore be straightforwardly adapted and only need a brief recollection here. 

At order $\alpha^6$, WBF $\llnnp$ production in association with two jets
proceeds via the scattering of two (anti-)quarks by $t$-channel
exchange of a weak gauge boson with subsequent emission of two $W^+$ bosons,
which in turn decay leptonically. Non-resonant diagrams where leptons are
produced via weak interactions in the $t$-channel also have been
included. For brevity, we will refer to $pp\to\llnnp jj$ as ``WBF $\wpp jj$''
production in the following, even though the electroweak production process
includes non-resonant diagrams that do not stem from a $W\to \ell\nu$
decay.

\begin{figure}
\centerline{ 
\epsfig{figure=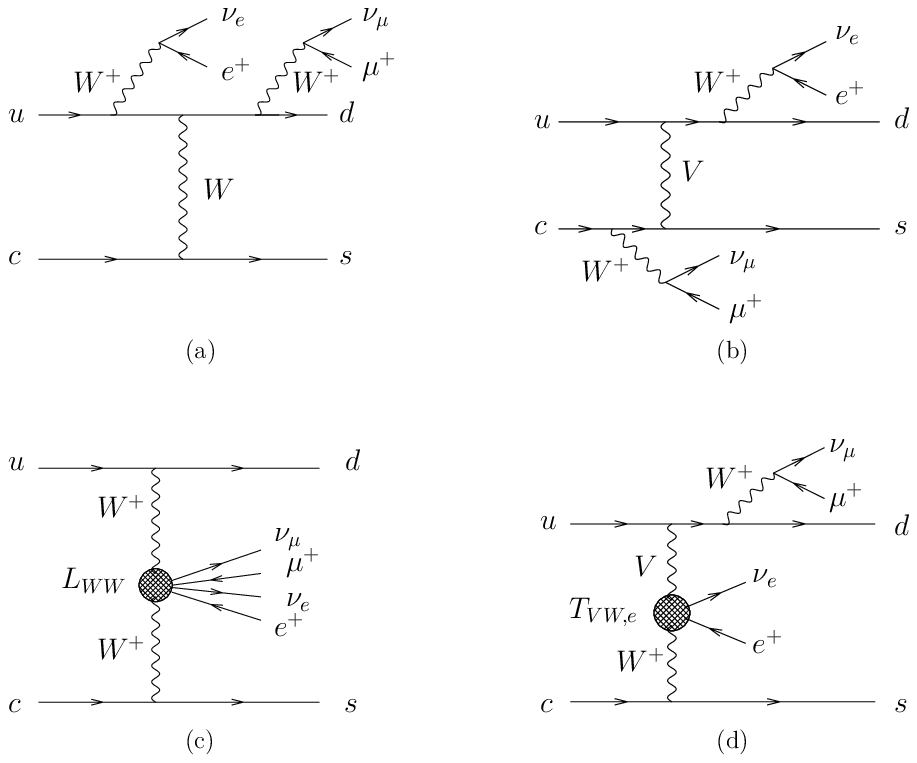,width=\textwidth,bb=80 345 560 745, clip=}
} 
\caption
{\label{fig:top} 
Feynman-graph topologies contributing to the Born process $uc\to d s \llnnp$. Not shown are diagrams analogous to (a) and (d) with $W^+$ emission from the lower quark line. $V$ denotes a $Z$ boson or a photon.  
}
\end{figure}

For each partonic sub-process, the 93 contributing Feynman diagrams can be grouped into four topologies, which are depicted for the representative 
$uc\to ds\llnnp$ mode in Fig.~\ref{fig:top}.  
%
%
%
%
The first two topologies correspond to the emission of two weak bosons from
the same~(a) or different~(b) quark lines with the $W^+$ decaying
leptonically. Topologies~(c) and~(d) comprise the so-called ``leptonic
tensors'' $L_{WW}$ and $T_{VW,e}$, which describe the tree-level amplitudes
of the sub-processes $W^+W^+\to\llnnp$ and $VW^+\to e^+\nu_e$, respectively.
In each case, $V$ stands for a virtual photon or $Z$ boson.  Not shown are
diagrams analogous to (a) and (d) with $W^+$ emission from the lower quark
line rather than the upper one and graphs which are obtained by permuting the
final-state leptons.

We disregard graphs with a weak-boson exchange in the $s$-channel with subsequent decay into a pair of jets, as they are strongly suppressed in the regions where WBF can be observed experimentally. For sub-processes with identical-flavor combinations such as $uu\to dd\llnnp$, in addition to the $t$-channel contributions discussed above also $u$-channel diagrams and their interference with the $t$-channel graphs arise. While we do take into account both $t$- and $u$-channel contributions, their interference cross section is kinematically strongly suppressed~\cite{Ciccolini:2007ec} and therefore not considered in the following. 

The relevant diagrams are combined in an efficient way, avoiding multiple
evaluation of recurring building blocks, and numerically evaluated using the
amplitude techniques of Refs.~\cite{Hagiwara:1985yu,Hagiwara:1988pp}.

For the NLO-QCD calculation, real emission and virtual corrections have to be
considered. Infrared singularities are handled via the subtraction formalism
in the form proposed by Catani and Seymour~\cite{Catani:1996vz}.

The real-emission contributions are obtained by attaching one extra gluon to
the tree-level diagrams sketched above. In addition to (anti-)quark initiated
sub-processes such as $uc\to dsg\llnnp$, contributions with a gluon in the
initial state (e.g.\ $ug\to ds\bar c\llnnp$) emerge. 

The virtual corrections comprise the interference of one-loop diagrams with
the Born amplitude. They are calculated in $d=4-2\epsilon$ dimensions in the
dimensional reduction scheme. Due to color conservation, contributions from
graphs with the virtual gluon being exchanged between the upper and the lower
quark line vanish at order $\alpha_s$, within our approximations. Only
self-energy, triangle, box and pentagon corrections to either the upper or
the lower quark line have to be considered. The singularities stemming from
infrared divergent configurations are calculated analytically and canceled by
the respective poles in the integrated counter-terms of the dipole
subtraction approach.
The finite pieces of the loop diagrams can be calculated in $d=4$
dimensions. The emerging two-, three-, and four-point tensor integrals are
evaluated numerically by a Passarino-Veltman reduction procedure. In order to
avoid numerical instabilities, for the computation of the pentagon integrals
we resort to the reduction scheme of Refs.~\cite{Denner:2002ii,Denner:2005nn},
which has already been employed in~\cite{Bozzi:2007ur,Campanario:2008yg}. We
monitor the numerical stability by checking Ward identities at every
phase-space point and find that the fraction of events which violate
electroweak gauge invariance by more than 10\% is at the permille level.  The
respective phase-space points are disregarded for the calculation of the
finite parts of the pentagon contributions and the remaining pentagon part is
compensated for this loss. As the pentagon contributions amount to only about
3$\permil$ of the full NLO result, the error induced by this approximation on
cross sections and distributions is negligible.

In order to ensure the reliability of our calculation, in addition to the
stability tests for the pentagon contributions, we performed the following
checks:
\begin{itemize}
\item 
We compared our tree-level and real-emission amplitudes for WBF $\llnnp jj$
scattering and crossing-related processes to the corresponding expressions
provided by the automatized matrix element generator
MadGraph~\cite{Alwall:2007st} and found full agreement within the numerical
accuracy of our program.
\item 
We compared our integrated cross section for $pp\to \llnnp jj$ via WBF within
the approximations discussed above to the full cross section provided by
MadEvent~\cite{Alwall:2007st} 
within typical WBF cuts. Our predictions fully agree with the corresponding
MadEvent results within the statistical errors of the two
calculations. Agreement between the two programs is also found for
representative distributions.
\end{itemize}


\section{\label{sec:num} Numerical Results}
The cross-section contributions discussed above have been implemented in a
fully flexible Monte-Carlo program, structured analogous to the {\tt VBFNLO}
code~\cite{Arnold:2008rz}.  In order to facilitate a comparison of the general
features of this production channel, similar settings and selection cuts as in
Refs.~\cite{Jager:2006zc,Jager:2006cp,Bozzi:2007ur} are employed.
We use the CTEQ6M parton distributions~\cite{Pumplin:2002vw} with
$\alpha_s(m_Z) = 0.118$ at NLO, and the CTEQ6L1 set at LO. Since in our
calculation quark masses are neglected, we entirely disregard contributions
from external $b$ and $t$ quarks. As electroweak input parameters, we have
chosen $m_Z=91.188$~GeV, $m_W=80.423$~GeV, and $G_F=1.166\times
10^{-5}/$~GeV$^2$. The other parameters, $\alpha_\mr{QED}$ and
$\sin^2\theta_W$, are computed thereof via tree-level electroweak
relations. For the reconstruction of jets from final-state partons we use the
$k_T$ algorithm~\cite{Catani:1992zp,Catani:1993hr,Ellis:1993tq,Blazey:2000qt}
with resolution parameter $D=0.7$. All our calculations are performed for a
$pp$~collider with center-of-mass energy of $\sqrt{s}=14$~TeV.

In order to clearly separate the WBF signal from various QCD backgrounds, we employ a set of dedicated selection cuts. We require at least two hard jets with
\be
\label{eq:ptjet-cut}
p_{Tj}\geq 20~\mr{GeV}\,,\quad
|y_j|\leq 4.5\,,
\ee
where $p_{Tj}$ denotes the transverse momentum and $y_j$ the rapidity of a
jet $j$ with the latter being reconstructed from massless partons of
pseudo-rapidity $|\eta_j|< 5$. The two jets of highest transverse
momentum are referred to as ``tagging jets''. We impose a large rapidity
separation between the two tagging jets,
\be
\Delta y_{jj} = |y_{j1}-y_{j2}|>4\,,
\ee
and furthermore demand that they be located in opposite hemispheres of the detector, 
\be
y_{j1}\times y_{j2}<0\,,
\ee
with an invariant mass
\be
M_{jj}>600~\mr{GeV}\,.
\ee

For our phenomenological analysis we focus on the $\llnnp$ leptonic final
state. Multiplying our predictions by a factor of two, results for the
four-lepton final state with any combination of electrons and/or muons and
the associated neutrinos (i.e., $e^+\nu_e \mu^+\nu_\mu$, $e^+\nu_e e^+\nu_e$,
and $\mu^+\nu_\mu \mu^+\nu_\mu$) can be obtained, apart from numerically
small identical-lepton interference effects.

To ensure that the charged leptons are well observable, we impose the lepton cuts
\bea
&p_{T\ell}\geq 20~\mr{GeV}\,,\qquad
|\eta_\ell|\leq 2.5\,,
&
\\
\label{eq:rl-cuts}
&\Delta R_{j\ell}\geq 0.4\,,\qquad
\Delta R_{\ell\ell}\geq 0.1\,,
&
\eea
where $\Delta R_{j\ell}$ and $\Delta R_{\ell\ell}$ denote the jet-lepton and lepton-lepton separation in the rapidity-azimuthal angle plane. 
%

\begin{figure}[!tb] 
\centerline{ 
\epsfig{figure=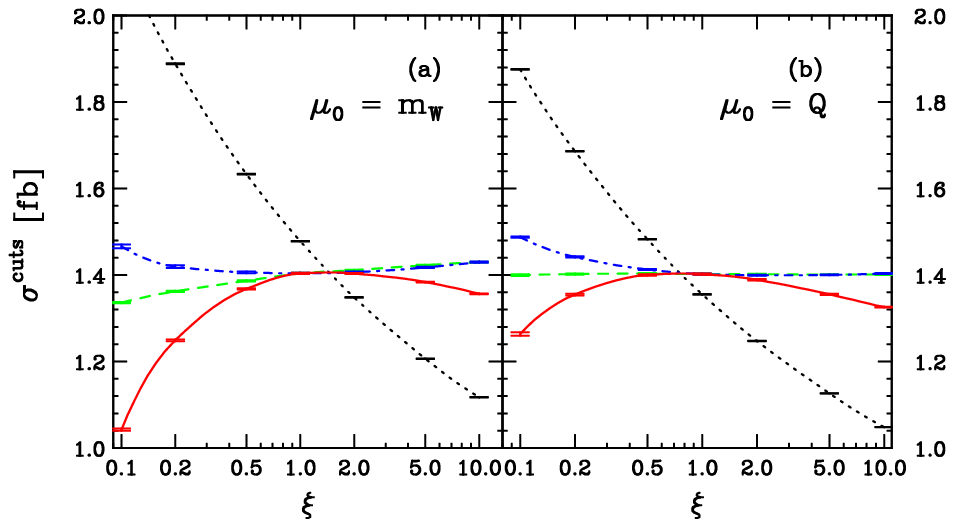,width=0.95\textwidth,clip=}
} 
\caption
{\label{fig:scale-dep} 
Dependence of the total WBF $pp\to \llnnp jj$  cross section at
the LHC on the factorization and renormalization scales for the two different
choices of $\mu_0$: $\mu_0=m_W$ and $\mu_0=Q$. 
The NLO curves show $\sigma^\mr{cuts}$ as a function of the scale parameter
$\xi$ for three different cases: $\mu_R=\mu_F=\xi\mu_0$ (solid red),
$\mu_F=\xi\mu_0$ and $\mu_R=\mu_0$ (dot-dashed blue), $\mu_R=\xi\mu_0$ and
$\mu_F=\mu_0$ (dashed green). The LO cross sections depend only on $\mu_F$
(dotted black). 
}
\end{figure}

As discussed in some detail in Ref.~\cite{Bozzi:2007ur}, a proper choice of
factorization and renormalization scales, $\mu_F$ and $\mu_R$, can help to
minimize the impact of higher order corrections on cross sections and
distributions in WBF reactions.  In the $W^\pm Z jj$ case, the momentum
transfer~$Q$ between an incoming and an outgoing parton along a fermion line
has been found to be more suitable than a constant mass scale. A similar
behavior can be observed in the $\wpp jj$ case, as illustrated by
Fig.~\ref{fig:scale-dep},
%
%
%
%
%
where we show the total cross section within the cuts of
Eqs.~(\ref{eq:ptjet-cut})-(\ref{eq:rl-cuts}), $\sigma^\mr{cuts}$, as
function of $\mu_F$ and $\mu_R$, which are taken as multiples of the scale
parameter $\mu_0$,
\be
\mu_F = \xi_F\mu_0\,,\qquad 
\mu_R = \xi_R\mu_0\,.
\ee
We distinguish between $\mu_0=m_W$ and $\mu_0=Q$ and vary $\xi_F$ and $\xi_R$ in the range $0.1$~to~$10$. At LO, the variation of the cross section with $\xi_F$ resembles the scale dependence of the quark distribution functions, $f_q(x,\mu_F)$, at relatively large values of $x$. Beyond LO, $\sigma^\mr{cuts}$ barely depends on $\mu_F$. The renormalization scale enters only at NLO via the strong coupling $\alpha_s(\mu_R)$. Due to the small size of the radiative corrections, the resulting $\mu_R$ dependence of the NLO cross sections is marginal. 
In Tab.~\ref{tab:sigma_cuts} 
%
%
\begin{table}[!ht]
\begin{center}
\begin{tabular}{|c|c|c|c|c|}
\hline
PDF set  &$\sigma^\mr{cuts}_\mr{LO}(\mu_0=m_W)$&$\sigma^\mr{cuts}_\mr{NLO}(\mu_0=m_W)$
	 &$\sigma^\mr{cuts}_\mr{LO}(\mu_0=Q)$  &$\sigma^\mr{cuts}_\mr{NLO}(\mu_0=Q)$  
	 \\
	 \hline
CTEQ6 &$1.478$~fb&$1.404$~fb&$1.355$~fb&$1.402$~fb 
	\\
	\hline
MSTW08 &$1.409$~fb&$1.372$~fb&$1.287$~fb&$1.369$~fb
	\\
	\hline
\end{tabular}
\caption{
\label{tab:sigma_cuts}
Cross sections for $pp\to\llnnp jj$ via WBF at 
LO and NLO within the cuts of Eqs.~(\ref{eq:ptjet-cut}-\ref{eq:rl-cuts}) for the scale
choices $\mu_F=\mu_R=m_W$ and $\mu_F=\mu_R=Q$ and two different sets of parton distribution functions. The statistical errors of the quoted results are at
the sub-permille level and therefore not given explicitly.  
}	
\vspace*{-.5cm}
\end{center}
\end{table}
%
%
%
we list the results for $\sigma^\mr{cuts}$ at LO and NLO for the two scale choices $\mu_0=m_W$ and $\mu_0=Q$ with $\xi_R=\xi_F=1$ and the settings defined above. For reference, we also quote the respective results obtained with the new MSTW08 parton distribution functions of Ref.~\cite{Martin:2009iq}. It is interesting to note that these results differ from the corresponding cross sections obtained with the CTEQ6 sets by about 5\% at LO and $2$~to~$3$\% at NLO. In the following we will stick to $\mu_F = \mu_R = Q$ and the CTEQ6 parton distributions. 

\begin{figure}[!tb] 
\centerline{ 
\epsfig{figure=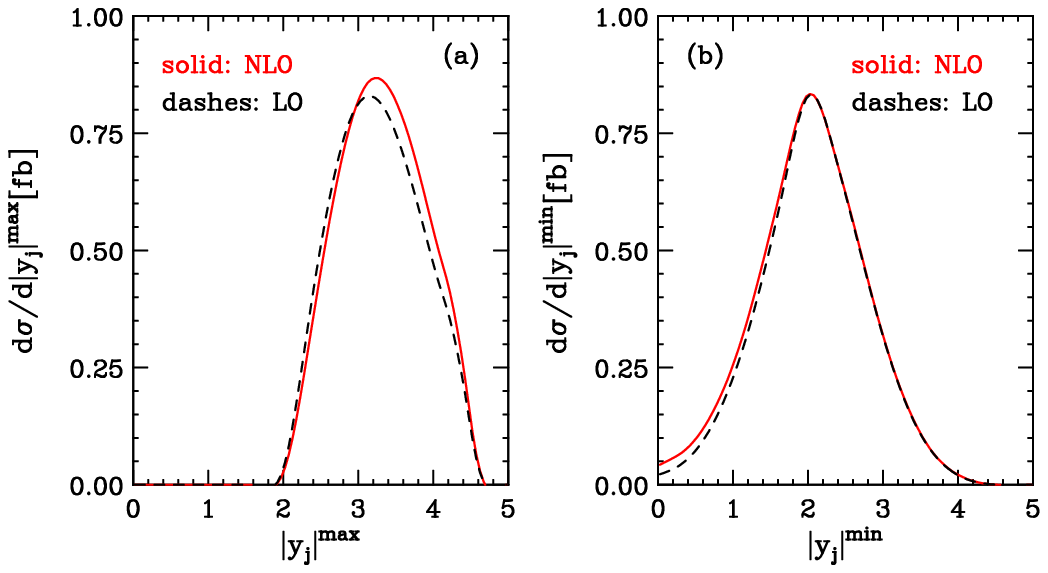,width=0.95\textwidth,clip=}
} 
\caption
{\label{fig:yjet} Rapidity distribution of the jet with the highest (a) and
with the lowest (b) value of the absolute value of the rapidity $|y_j|$ in WBF
$pp\to \llnnp jj$ production at the LHC at LO (dashed black line) and NLO (solid red
line).  }
\end{figure}

A very characteristic feature of WBF-induced processes is the distribution of the tagging jets, which are located in the far-forward and -backward regions of the detector. Due to the color-singlet nature of the $t$-channel weak boson exchange, the central-rapidity range of the detector exhibits little jet activity. As this feature is an important tool for the suppression of QCD backgrounds which typically exhibit rich hadronic activity at low rapidities~\cite{Rainwater:1999sd}, 
it is important to estimate the impact of NLO-QCD corrections on the rapidity distributions of the observed jets. To this end, in Fig.~\ref{fig:yjet}~(a)  
%
%
%
%
we display the distribution of the jet with the highest absolute value of
rapidity at LO and NLO.  The respective results for the jet with the lowest
absolute value of rapidity are shown in Fig.~\ref{fig:yjet}~(b).
While at LO each final-state parton corresponds to a tagging jet, at NLO two or three jets may arise due to the extra parton in the real-emission contributions. 
Because of this extra parton, the probability to find a jet of very high or very low rapidity is larger at NLO than at LO.

\begin{figure}[!tb] 
\centerline{ 
\epsfig{figure=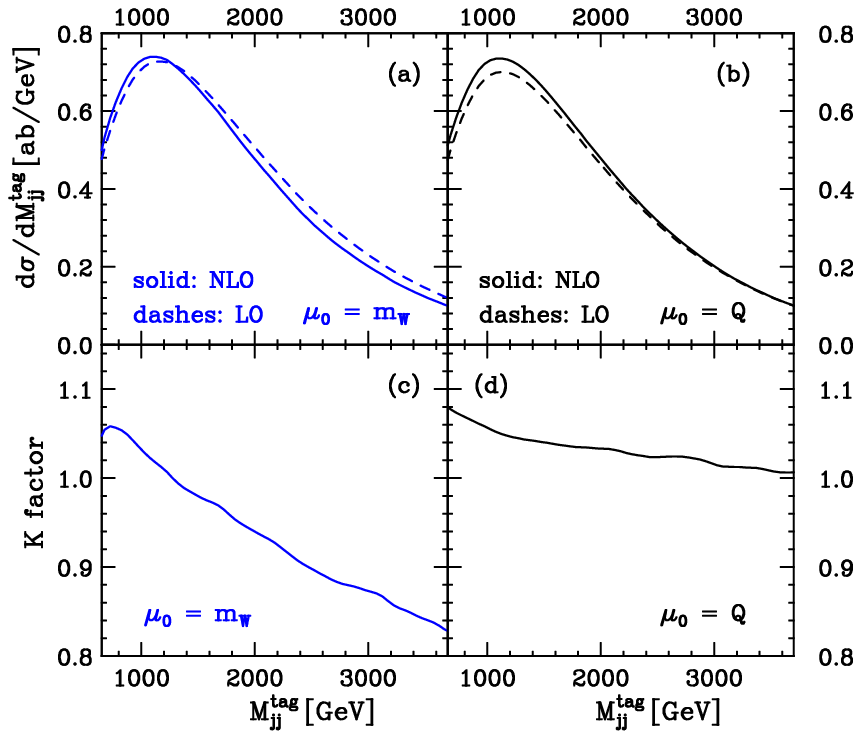,width=0.95\textwidth,clip=}
} 
\caption
{\label{fig:mjj} 
Invariant mass distribution of the tagging jets in WBF $pp\to \llnnp jj$ production at the LHC for two different choices of $\mu_0$ [panels (a) and (b)] at 
LO (dashed) and NLO (solid). Their ratios, the $K$ factors as defined in Eq.~(\ref{eq:kfac}), are displayed for $\mu_0 = m_W$ in panel~(c) and for  $\mu_0 = Q$ in panel (d).
}
\end{figure}

In Fig.~\ref{fig:mjj}, 
%
%
%
%
%
the invariant mass distribution of the two tagging jets is shown for two choices of $\mu_0$ together with the dynamical $K$ factor, which is defined via
\be
\label{eq:kfac}
K(M_{jj}) = \frac{d\sigma_\mr{NLO}/dM_{jj}}{d\sigma_\mr{LO}/dM_{jj}}\,.
\ee
The $M_{jj}$ distribution was scrutinized in Ref.~\cite{Bozzi:2007ur} to
demonstrate the impact of the choice of factorization and renormalization
scales on the perturbative stability of calculations for WBF processes. Using
the example of WBF $W^+Zjj$ production it was shown that small NLO-QCD
corrections to $d\sigma/dM_{jj}$ are obtained, if $\mu_F$ and $\mu_R$ are
identified with $Q$. A similar behavior is observed for the $W^+W^+jj$ case,
with a $K$~factor close to one over the entire range of $M_{jj}$ for
$\mu_0=Q$, but pronounced shape distortions at LO for $\mu_0=m_W$.


\section{\label{sec:conc} Summary and Conclusions}
In this work, we have presented a NLO-QCD calculation for WBF $\wpp jj$ and
$\wmm jj$ production at the LHC, which takes leptonic decays of the weak
gauge bosons fully into account. We have developed a flexible Monte-Carlo
program that allows for the calculation of cross sections and distributions
within typical WBF cuts. The QCD corrections to the integrated cross section
are modest, changing the leading-order results by less than about 7\%. The
residual scale uncertainties of the NLO predictions are at the 2.5\% level,
indicating that the perturbative calculation is under excellent control. In
accordance with our earlier work on WBF $W^\pm Zjj$ production we found that
size and shape of the NLO results can be best approximated by the
corresponding LO expressions, if the momentum transfer, $Q$, of the
$t$-channel electroweak boson is chosen as factorization scale.


\begin{acknowledgments}
We are grateful to Malgorzata Worek for valuable discussions and to Michel Herquet and Rikkert Frederix for useful comments on the performance of MadEvent for high-multiplicity processes. 
We would like to thank Stefan Kallweit for pointing out to us a typo in the list of cuts we have used, in the first version of the paper.  
This work was supported by the Initiative and Networking Fund of the Helmholtz
Association, contract HA-101 ("Physics at the Terascale").
D.~Z.\ was supported by the Deutsche Forschungsgemeinschaft under SFB TR-9 ``Computergest\"utzte Theoretische Teilchenphysik''.

\end{acknowledgments}


\bibliography{papers}

\end{document}